\documentclass[aps,prd,amsmath,amsfonts,a4paper,11pt,reprint,twocolumn,square,numbers,showpacs,superscriptaddress,floatfix,sort&compress,nofootinbib]{revtex4-1}
\usepackage{hyperref}
\usepackage{amssymb}
\usepackage{bm}
\usepackage{natbib}
\usepackage{cancel}
\usepackage{graphicx}
\expandafter\let\csname equation*\endcsname\relax
\expandafter\let\csname endequation*\endcsname\relax
\usepackage{amsmath}
\usepackage{color}

\RequirePackage{ifpdf}
\ifpdf
\else % ordinary latex seems to include these as eps files without a problem
\fi

\newcommand{\para}[1]{\par\vspace{2mm}\noindent\textbf{\emph{{#1}}.---}}

\usepackage{titlesec}
\titlespacing*{\section}{0pt}{0.6cm}{0.5cm}

\newcommand{\beq}{\begin{equation}}
\newcommand{\eeq}{\end{equation}}

\newcommand{\bea}{\begin{eqnarray}}
\newcommand{\eea}{\end{eqnarray}}

\begin{document}

\begin{center}
\rightline{\small DESY-18-165}
\vskip -3cm
\end{center}

	\title{On uplifts by warped anti-D3-branes}

	\author{Jakob Moritz}
	\email{jakob.moritz@desy.de}
	\author{Ander Retolaza}
	\email{ander.retolaza@desy.de}
	\author{Alexander Westphal}
	\email{alexander.westphal@desy.de}
	\affiliation{Deutsches Elektronen-Synchrotron, DESY, Notkestra\ss e 85, 22607 Hamburg, Germany}

	\begin{abstract}
		
	In this note we outline the arguments against the ten-dimensional consistency of the simplest types of KKLT de Sitter vacua, as given in~\cite{Moritz:2017xto}. We comment on parametrization proposals within four-dimensional supergravity and express our disagreement with the recent criticism by the authors of~\cite{Kallosh:2018wme}.

	\end{abstract}
	
	\maketitle

 	\section{Introduction}
\label{intro}

%%%%%%%%%%%%%%%%%%%%%%%%%%%%%%%%%%%%%%%%%%%%%%%%%%%%%%%%%%%%%%%%%%%%%%

Whether or not string theory has semi-classical solutions with the isometries of four dimensional de Sitter space is the subject of an ongoing debate. The renewed interest concerning this question is partially due to a very recent conjecture that states that at each point in scalar field space,
the gradient of the scalar potential is bounded from below by the vacuum energy (in Planck units)~\cite{Obied:2018sgi}. If true, positive energy solutions must always be time-dependent due to the rolling of scalar fields.\footnote{As stated, the conjecture~\cite{Obied:2018sgi} would even forbid any de Sitter critical point. This is hard to justify both from a stringy perspective \cite{Roupec:2018mbn,Conlon:2018eyr} as well as from the bottom-up \cite{Denef:2018etk,Murayama:2018lie,Choi:2018rze}. However, it is possible that a statement could hold that is similar in spirit, allows for de Sitter critical points but not minima (see e.g. \cite{Garg:2018reu,Conlon:2018eyr}).}  This conjecture is fueled partly by classical no-go theorems called Maldacena-N\'{u}\~{n}ez theorems~\cite{deWit:1986mwo,Maldacena:2000mw} against four dimensional de Sitter critical points in ten and eleven-dimensional supergravity. These theorems apply whenever each singularity that occurs in the compactification is controlled by a stress-energy tensor that satisfies a certain energy condition. However, these theorems can in principle be evaded once either 
\begin{itemize}
	\item sources that violate the energy condition are implemented in a controlled way or
	\item  genuinely four dimensional (e.g. non-perturbative) effects play an important role that \textit{cannot} be treated semi-classically within a higher dimensional framework.
\end{itemize}
One of the most popular mechanisms for generating $4D$ de Sitter vacua (the KKLT mechanism~\cite{Kachru:2003aw}) has been argued to operate along the second option. Since this mechanism can be considered an important counter example to the conjecture~\cite{Obied:2018sgi} it is clearly interesting to put it under scrutiny, either to improve our understanding of the mechanism or to find good reasons to reject it. It is important to carefully weigh both supporting and critical arguments (even though problems with any one de Sitter construction are not the end of the de Sitter landscape): 
on the one hand being able to reject KKLT would supply non-trivial evidence supporting the absence of de Sitter vacua in string theory and, if the conjecture is proven, it would provide an unprecedented concrete and experimentally testable prediction.

Last year we came to the conclusion that the simplest example of the KKLT mechanism cannot produce four dimensional de Sitter vacua because this would violate the ten dimensional Maldacena-N\'{u}\~{n}ez consistency conditions. Since this conclusion has been criticized recently from various angles we would like to take the opportunity to explain which types of criticism in our opinion can form a basis for a fruitful discussion and which do not. To this end we will shortly review the main features of KKLT, our objection to it, and comment on recent criticism.

\section{KKLT in a nutshell}
KKLT work within the landscape of Calabi-Yau orientifolds of type IIB string theory (in its supergravity approximation) with three-form flux quanta turned on. Within this setup it is known that a subset of the light scalar degrees of freedom obtain a large mass (at least as long as fluxes are dilute). These are the axio-dilaton as well as the complex structure moduli. After integrating out these degrees of freedom, we are left with an effective supergravity model for the K\"ahler moduli. If there is only a single K\"ahler modulus $T$, the model is parametrized by a K\"ahler potential and superpotential
\begin{equation}
K=-3\log(T+\overline{T}) +const\, ,\quad W=W_0=const.
\end{equation}
The constant $W_0$ comes from evaluating the Gukov-Vafa-Witten (GVW) flux superpotential at the stabilized value of the complex structure moduli and the axio-dilaton. Appropriate choices of flux quanta allow us to tune $|W_0|$ extremely small. The model is of \textit{no-scale} type\footnote{See however \cite{Sethi:2017phn}. 
} and its scalar potential vanishes so the K\"ahler modulus $T$ remains massless.

KKLT proposed to include non-perturbative effects such as gaugino condensation on a stack of $N$ $D7$ branes which would contribute a superpotential term $Ae^{-aT}$, with $A=\mathcal{O}(1)$ and $a=\frac{2\pi}{N}$. It was shown that the quantum corrected superpotential leads to an Anti-de-Sitter vacuum with K\"ahler modulus stabilized at $T_0\approx \frac{1}{a}\log\left[-\frac{A}{3W_0}\log(|W_0|^{-2})\right]$ for $|W_0|\ll 1$. Once $W_0$ is sufficiently small, we can self-consistently neglect perturbative and higher non-perturbative corrections.

The next step consists of adding a SUSY breaking warped anti-D3-brane. Dimensionally reducing the anti-D3-brane on the no-scale GKP~\cite{Giddings:2001yu} flux background gives rise to a further term in the scalar potential~\cite{Kachru:2002gs}
\begin{equation}
V_{\overline{D3}}\sim \frac{2a_0^4 T_3}{(T+\overline{T})^2}\, ,
\end{equation}
where $a_0^2$ is of order the IR warp factor $\exp(2{\cal A}_0)$ of a Klebanov-Strassler throat and can easily take exponentially small values, and $T_3=\mathcal{O}(1)$ is the brane tension in Planck units. If this is correct, de Sitter vacua with tunable cosmological constant exist and the anti-brane potential does not interfere significantly with the stabilization mechanism. Alternatively, one may parametrize the \textit{uplift} with the help of a nilpotent chiral superfield $S$~\cite{Ferrara:2014kva,Kallosh:2014wsa,Bergshoeff:2015jxa,Kallosh:2015nia}, with K\"ahler and superpotential
\begin{equation}\label{eq:WK_KKLTUplift1}
K=-3\log(T+\bar{T}-S\bar{S})\, ,\quad W=W_0+Ae^{-aT}+bS\, ,
\end{equation}
where $b\sim a_0^2\sqrt{T_3}$, and $S$ is constrained by $S^2=0$.  

\section{The ten dimensional perspective}
In~\cite{Moritz:2017xto} we have analyzed the KKLT construction from a ten-dimensional point of view. Obviously, our conclusions are valid only if a ten-dimensional description of gaugino condensation on a seven-brane stack exists. In our opinion a reasonable objection can be that gaugino condensation is usually understood as a genuinely four-dimensional phenomenon. However, there exists compelling quantitative evidence that non-perturbatively stabilized KKLT vacua can be consistently described by plugging in a non-trivial expectation value for the gaugino bilinear $\langle\lambda\lambda\rangle$ into the seven-brane action and compute the ten-dimensional backreaction~\cite{Baumann:2010sx,Heidenreich:2010ad,Dymarsky:2010mf}. By doing so, one may for instance match the known scalar potential for mobile $D3$-brane moduli~\cite{Berg:2004ek,Baumann:2006th,Baumann:2010sx} that are induced by gaugino condensation. Moreover, the SUSY condition $0=D_T W\sim W_0+\#\langle\lambda\lambda\rangle$ can be rediscovered in ten dimensions~\cite{Moritz:2017xto}. To us the evidence is convincing enough to believe that a ten dimensional description of KKLT vacua in the spirit of~\cite{Baumann:2010sx,Heidenreich:2010ad,Dymarsky:2010mf} is consistent. In our opinion criticism of this point of view is perfectly acceptable but the highly non-trivial evidence for it should be taken into consideration carefully.

Under the above assumption one may estimate the effective energy momentum tensor that is induced by gaugino condensation and check whether or not it provides a loop-hole to the Maldacena-N\'{u}\~{n}ez theorem. For type IIB supergravity there exists a strengthened form of the theorem that can be derived from a tadpole constraint which takes the form~\cite{deAlwis:2003sn,Heidenreich:2010ad} (see e.g. also~\cite{Giddings:2001yu,Dasgupta:2014pma})
\begin{eqnarray}
0&=&\! \int\limits_{CY_3} \! \! \! d^6y\sqrt{g_6}\left[e^{6A}R_{4}+e^{8A}\dfrac{\Delta^{loc}}{2\pi}\right.\nonumber\\
&&\qquad\qquad\;\;\;\; \left.+e^{8A}\;(\text{positive semi-definite})\right]\, ,
\end{eqnarray} 
where \ $\Delta^{loc}\equiv\frac{1}{4}\left(T^m_m-T^{\mu}_{\mu}\right)^{\text{loc}}- T_3\rho_3^{\text{loc}}$ is a combination of the energy momentum tensor and D3-brane charge of localized objects, and $R_{4}$ is the Ricci-scalar of the four dimensional vacuum. For the classical no-scale solutions all terms of the integrand vanish individually and the tadpole bound is satisfied. Clearly, for a de Sitter vacuum $R_{4}>0$ so there must exist some localized object that has a negative $\Delta^{loc}$.

One can convince oneself that this is a problem for the KKLT construction: in the supersymmetric situation the only non-vanishing terms are sourced by the gaugino condensate. Since for a supersymmetric background $R_{4}\leq 0$, the sum of the other terms must give a positive contribution. To leading order in the (small) gaugino condensate all non-vanishing terms localize along the seven brane stack so can be encoded in an effective energy momentum tensor $T_{\lambda\lambda}$ that must satisfy $\Delta^{loc}_{\lambda\lambda}\sim |\lambda\lambda|^2>0$, at least in the supersymmetric situation.

In passing to the non-supersymmetric situation one has to include the localized energy momentum tensor of the anti-brane which by itself does not allow to evade the no-go theorem, i.e. it gives a positive contribution to $\Delta^{loc}$~\cite{Giddings:2001yu,Heidenreich:2010ad}. Hence, even after inclusion of the SUSY breaking anti-brane the no-go theorem \textit{still} applies\footnote{Strictly speaking one also has to estimate to what extend the presence of the anti-brane can change the value of the gaugino condensates contribution to $\Delta^{loc}_{\lambda\lambda}$. Under the assumption that the magnitude of the condensate is not reduced by a large amount (i.e. by a whole cycle-volume) one may argue that the only non-negligible backreaction can be that the volume modulus finds a new stabilized value and hence the value of the condensate adjusts. This is enough to conclude the above.}.

From this we have concluded that single modulus KKLT vacua cannot be lifted to de Sitter vacua via the inclusion of warped SUSY breaking sources such as anti-branes. We would like to emphasize that this was the main idea presented in~\cite{Moritz:2017xto}. One may \textit{either} criticize the ten-dimensional approach from the start (as was done for instance in~\cite{Cicoli:2018kdo}) or accept it and argue about the proper implementation within $4D$ supergravity.

\textit{If} one is convinced that our $10D$ conclusion is correct one may debate over how it can be implemented within a corrected $4D$ supergravity model. As a first step toward such a $4D$ parametrization we have proposed a simple model that does the job (in some regime) and we will comment on this and the criticism of~\cite{Kallosh:2018wme} in the next section.

\section{Four dimensional supergravity implementations}

\begin{figure}[t!]
	\begin{center}
		\includegraphics[keepaspectratio,width=8.5cm]{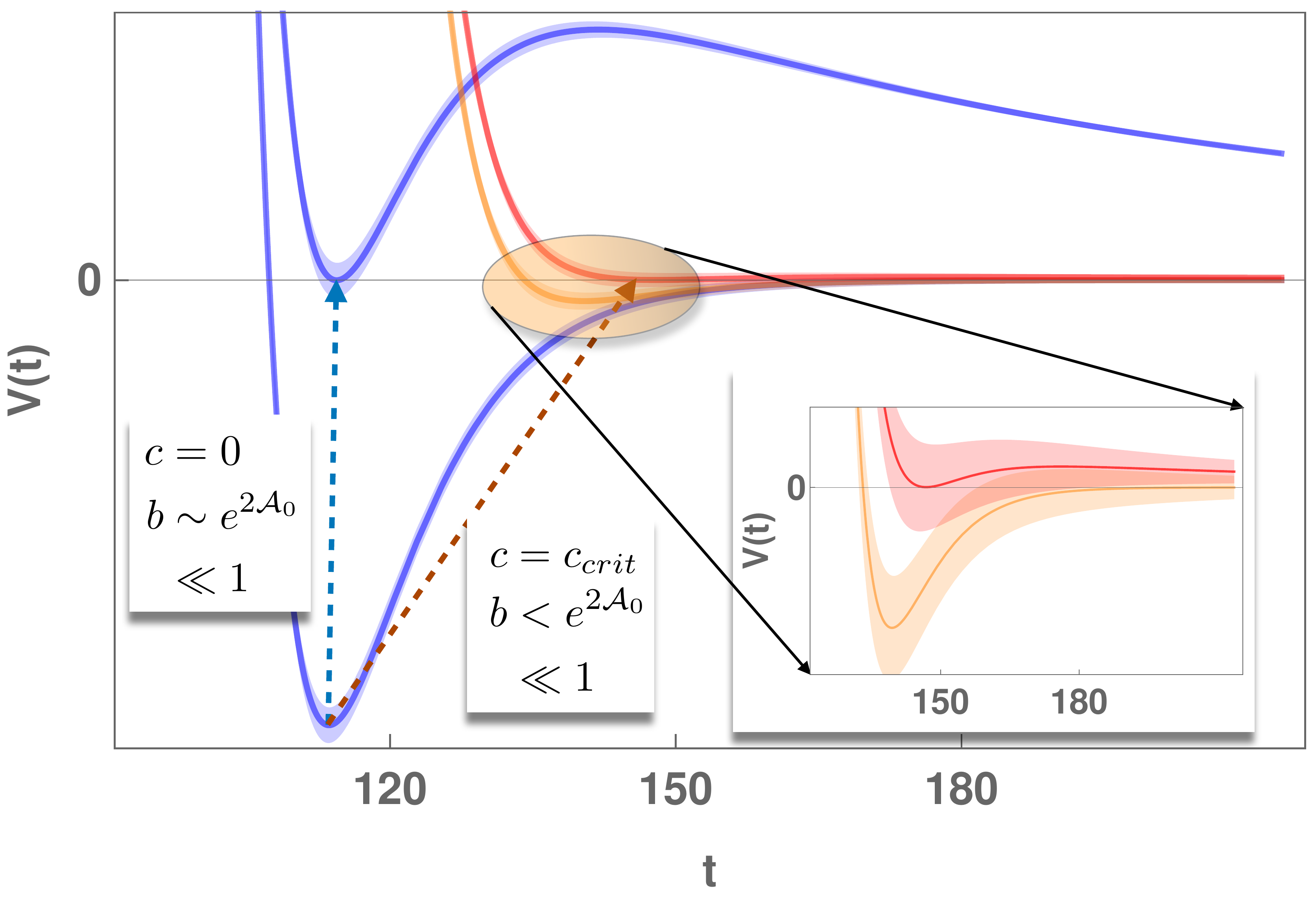}
	\end{center}\vspace*{-0.5cm}
	\caption{In blue: the supersymmetric KKLT potential with an uplift by $c=0$, $b\sim e^{2{\cal A}_0}$ in blue error bars in orange (assuming $p=\frac{1}{2}$). Perturbative corrections to the potential are negligible. In orange: partially uplifted scalar potential in the limit of marginally controlled uplift from the $c$-coupling at $c= c_{crit}$ (and $\gamma\approx 6$), $b=0$. Perturbative corrections can no longer be neglected. In red: uplifted scalar potential in the limit of marginally controlled uplift from the $c$-coupling at $c=c_{crit}$ , and additional $b$-coupling uplift with $0<b<e^{2{\cal A}_0}$. Again, perturbative corrections can not be neglected, and putative dS minima in this regime are not trustworthy. In all cases we have chosen $A=1$, $a=0.1$, $W_0=-10^{-4}$.}
	\label{fig:errorbars}
\end{figure}

Based on  a ten dimensional analysis we have argued that within a consistent truncation to leading order in inverse cycle volumes there are no de Sitter KKLT vacua within a controlled regime. Of course, it would be desirable to understand this from the four dimensional perspective. In~\cite{Moritz:2017xto} the following interpretation\footnote{This interpretation is correct at least in certain $6d$ toy models that are similarly constrained by the Maldacena-N\'{u}\~{n}ez theorem.  } was given: 

\begin{align}\label{Statement}
&\textit{As the uplift potential (say the IR warp factor)}\nonumber\\
&\textit{is dialed up, the volume modulus is pushed to}\nonumber\\
&\textit{larger volumes in such a way that it is impossible}\\
&\textit{to reach positive vacuum energy (see Figure~\ref{fig:errorbars}).}\nonumber 
\end{align}

We have proposed to implement this behavior by modifying the $4D$ SUGRA as follows
\begin{equation}\label{eq:WflattenedUplift1}
W\longrightarrow W_0+Ae^{-aT}+c S e^{-aT}+ b S\, ,
\end{equation}
for some unsuppressed coefficient $c$. This proposal matches the outcome of our $10D$ arguments by realizing the above interpretation~\eqref{Statement}. For this to happen the dominant part of uplifting must arise from the $c$-coupling, and \emph{not} from the $b$-coupling. In particular, only choosing a \textit{large enough} coefficient $c$ (where `large enough' will be made more precise below) provides a match to the $10D$ results.

One may convince oneself that for $b=0$ there are no de Sitter vacua and the coupling proportional to $c$ mediates a large back-reaction on the K\"ahler modulus $T$ as the vacuum energy increases. For any given value of $c$ we can turn on the small coefficient $b$ until eventually one reaches positive vacuum energy and the \textit{additional} backreaction on $T$ that comes from turning on $b\neq 0$ is small. However, if $c$ is \textit{large enough} to provide the dominant part of uplifting to zero vacuum energy, such de Sitter ``vacua" cannot be trusted within a truncation to the leading order K\"ahler potential. This is because contributions to the scalar potential from perturbative corrections to the K\"ahler potential can no longer be argued to be negligible. The model thus matches our $10D$ result within the margin of theoretical error if $c$ is large enough in the sense we now describe.

The $T$-dependent scalar potential around the supersymmetric AdS KKLT vacuum ($b=c=0$) looks rather similar to the potential around the non-supersymmetric AdS point with $c\neq 0$. Naively one might believe that both vacua are equally well controlled. However, perturbative corrections to the scalar potential in the form of volume suppressed $\alpha'$ corrections are expected to take the form
\begin{equation}
\delta V\sim e^{K_0}|W_0|^2\times \frac{1}{(T+\bar{T})^p}\, ,\quad p>0\, ,
\end{equation}  
where $K_0$ is the tree level K\"ahler potential and $p=1/2,1,3/2$ for corrections arising at ${\cal O}(\alpha'),{\cal O}(\alpha'^2),{\cal O}(\alpha'^3)$, respectively.

The vacuum energy of the supersymmetric vacuum is given by $V_{\text{SUSY}}=-3e^{K}|W|^2\approx -3e^{K_0}|W_0|^2$. Therefore, in the large volume regime one may neglect $\alpha'$ corrections to the scalar potential expanded around the SUSY vacuum. In contrast, the vacuum energy of the non-supersymmetric vacuum is parametrized by the value of $c$. If $c$ is large enough, one has $|V_{\text{SUSY}}|\gg |V_{\cancel{\text{SUSY}}}|\sim \delta V$ and perturbative corrections start to give important contributions to the scalar potential. We plot both scalar potentials with their margins of theoretical error in Figure \ref{fig:errorbars}.

Hence, there is a critical value of $c$,
\begin{equation}\label{eq:UpliftControlLimit1}
c_{crit}\equiv  \gamma A \sqrt{a\log(|A/W_0|)}\, ,
\end{equation}
with numerical coefficient $\gamma$, such that
\begin{itemize}
\item for $c<c_{crit}$ using the $b$-coupling to provide the missing uplift to zero vacuum energy contradicts our $10D$ outcome~\eqref{Statement},
\item while for $c\gtrsim c_{crit}$ any dS minima created by adding the $b$-coupling are in the regime $|V_{\text{SUSY}}|\gg |V_{\cancel{\text{SUSY}}}|\sim \delta V$ where the scalar potential cannot be reliably predicted.
\end{itemize}
In the ``counter example" given in~\cite{Kallosh:2018wme} the authors have chosen $c=1$ which is simply not large enough in the sense given above -- i.e. for $c=1$ most of the uplift to zero vacuum energy still comes from the b-coupling.

A further criticism that was spelled out~\cite{Kallosh:2018wme} is that the model cannot be valid on all of moduli space because at points where $c e^{-aT}+b=0$ supersymmetry is restored. Certainly we would not claim that the model is a good description in this regime. 

The model was only designed to illustrate the effect of unsuppressed exponential couplings so in particular we did not claim that it is a unique consequence of our ten dimensional analysis. In fact the model can be easily generalized to a whole class of models that all exhibit the effect that we are after  while avoiding the problem  just mentioned. One simply starts with the superpotential of eq.~\eqref{eq:WK_KKLTUplift1}, transforms the classical warp factor $b$ into the K\"ahler potential by a field redefinition of $S$ as in~\cite{GarciadelMoral:2017vnz},
\begin{equation}
K=-3\ln\left(T+\bar T-\frac{S\bar S}{b^2}\right)
\end{equation}
and then replaces
\begin{equation} \label{eq:quantum-warp}
b^2\rightarrow  b^2+b(f(T+\bar{T})e^{-a T}+c.c.)+g(T+\bar{T})e^{-2a\text{Re}T}\, ,
\end{equation}
with some power law functions $f$ and $g$. For the special choice $f=\bar{c}\in \mathbb{C}$, $g=|c|^2$  we obtain the simple parametrization that was originally proposed. 

However, we may instead choose $g(T+\bar{T})=g_1\cdot (T+\bar{T})$, with $g_1\in \mathbb{R}_+$. One can check that the bound analogous to \eqref{eq:UpliftControlLimit1} reads  
\begin{equation}
g_1\gtrsim g_{crit} \equiv \gamma' a^2A^2\, ,
\end{equation}
again for some numerical constant $\gamma'$.

Clearly, the problem pointed out in~\cite{Kallosh:2018wme} has disappeared because the expression in  (\ref{eq:quantum-warp}) does not have a root on the positive half plane $T+\bar{T}>0$. The model is ``better" than the simplest model we started with for a number of reasons: first, one can match the ten dimensional result with a simple bound that does not scale with $W_0$ as awkwardly as the one of eq.~\eqref{eq:UpliftControlLimit1}. Second, the quantity in (\ref{eq:quantum-warp}) can be interpreted as a quantum corrected IR warp factor very much in the spirit of~\cite{Giddings:2005ff}.

\section{Conclusions}

In this note we have attempted to succinctly review our arguments in~\cite{Moritz:2017xto} which show that the most simple setup exemplifying the KKLT mechanism where the K\"{a}hler modulus is stabilized by a single gaugino condensate does not evade the Maldacena-N\'{u}\~{n}ez theorem. As a consequence, the warped anti-D3-brane fails to uplift the vacuum energy to positive values. After summarizing the $10D$ argument we have reviewed the $4D$ parametrization proposals made in~\cite{Moritz:2017xto} to meet the $10D$ results. Finally we have expressed our disagreement with certain points made in ~\cite{Kallosh:2018wme}. Besides analyzing the $4D$ parametrization proposals, a discussion of the validity of our $10D$ argument is clearly necessary and must be carried out elsewhere.

\para{Acknowledgements} We are grateful to  Renata Kallosh, Andrei Linde and Marco Scalisi for many discussions and to Arthur Hebecker for useful comments on a previous version of the paper.  
 AR is grateful to the Institut de Physique Th\'{e}orique IPhT CEA Saclay and the Instituto de F\'{i}sica Te\'{o}rica IFT UAM/CSIC, Madrid for kind hospitality while part of this work was done. This work was supported by the ERC Consolidator Grant STRINGFLATION under the HORIZON 2020 grant agreement no.
647995.

\bibliography{refs}
\end{document}